\documentclass[article]{JHEP3}
\usepackage{amsmath,amssymb}
\usepackage{mathtools}
\usepackage{cite}
\usepackage{verbatim}
\usepackage{float}
\usepackage{subcaption}
\usepackage{amsfonts}
\usepackage[utf8]{inputenc}
\usepackage{graphicx}

\allowdisplaybreaks

\newcommand{\be}{\begin{equation}}
\newcommand{\ee}{\end{equation}}
\newcommand{\bea}{\begin{eqnarray}}
\newcommand{\eea}{\end{eqnarray}}
\newcommand{\bean}{\begin{eqnarray*}}
\newcommand{\eean}{\end{eqnarray*}}

\def\beq{\begin{equation}}

\def\eeq{\end{equation}}
\def\R{\mathcal{R}}

\def\Im{\mathop{\rm Im}}
\relax

\def\a{\alpha'}

\title{Greybody factors of large Gauss-Bonnet $d$-dimensional black holes}

\author{Filipe Moura$^a$ and Jo\~ao Rodrigues$^b$
\\
\\
$^{a}$
Departamento de Matem\'atica, Escola de Tecnologias e Arquitetura, \\ ISCTE - Instituto Universit\'ario de Lisboa \\ and Instituto de Telecomunica\c c\~oes,
\\Av. das For\c cas Armadas, 1649-026 Lisboa, Portugal\\
\email{fmoura@lx.it.pt}
\\
\\
$^{b}$
Centro de An\'alise Matem\'atica, Geometria e Sistemas Din\^amicos,\\ Departamento de Matem\'atica,\\ Instituto Superior T\'ecnico,\\
Av. Rovisco Pais, 1049-001 Lisboa, Portugal\\
\email{joao.carlos.rodrigues@tecnico.ulisboa.pt}
}

\abstract{We compute analytically greybody factors for $d$-dimensional large spherically symmetric black holes with Gauss-Bonnet corrections in the high frequency limit. Our calculations include both the eikonal limit, where the real part of the frequency of the scattered wave
is much larger than the imaginary part, and the highly damped case,where the imaginary part of the frequency is much larger than the real part, addressing the emission of gravitons and test scalar fields, and yielding full transmission and reflection scattering coefficients. We consider tensorial, vectorial and scalar gravitational perturbations, obtaining the same result for the three cases in the highly damped limit, but not in the eikonal limit.
}

\begin{document}



\vfill

\eject

\section{Introduction}
\noindent

In the famous calculation of the Hawking radiation~\cite{Hawking:1975vcx}, it was shown that black holes have a thermal spectrum. The expectation value for the number of particles emitted with a certain frequency $\omega$ is
\begin{equation}
    \Big\langle n(\omega) \Big\rangle = \frac{\gamma(\omega)}{e^{\frac{\omega}{T_{\text{H}}}}\pm 1}, \label{gfdef}
\end{equation}
where $T_{\text{H}}$ is the Hawking temperature of the black hole spacetime and the $+/-$ sign addresses radiation composed by fermions or bosons, respectively. The frequency-dependent factor $\gamma(\omega)$ is the greybody factor. Integrating the expression above over the entire frequency spectrum yields the black hole emission rate.

The greybody factor is directly connected to the asymptotic observation of Hawking radiation. Indeed, the actual spectrum observed by an asymptotic observer is different from a blackbody spectrum. Black hole greybody factors are functions characterizing the tunnelling probability of perturbations through the black hole effective potential~\cite{Parikh:1999mf}. A greybody factor is a transmission probability: a quantity that describes the deviation of the Hawking radiation from a pure blackbody radiation. The computation of these factors is therefore crucial in order to understand the Hawking radiation from a semiclassical point of view.

Black hole greybody factors are also relevant in the modelling of post-merger gravitational wave ringdown signals~\cite{Oshita:2022pkc,Oshita:2023cjz,Okabayashi:2024qbz,Konoplya:2024lir}. The ringdown phase, describing the relaxation of the remnant black hole, is described using quasinormal modes, corresponding to exponentially damped oscillations. Although black hole spectroscopy is based on quasinormal modes, these are known to be highly sensitive to relatively small deformations of the black hole geometry. Differently than the quasinormal spectra, greybody factors are stable under such small deformations~\cite{Rosato:2024arw,Oshita:2024fzf}. Therefore, greybody factors may be useful not only for computing the spectrum of Hawking radiation but also in the context of astrophysical observations of gravitational waves from black holes.

Greybody factors corresponding to minimally coupled massless scalar fields have been computed in asymptotically flat black holes \cite{Page:1976df,Page:1976ki,Starobinsky:1973aij} in $d=4$ Einstein gravity. More recently, these studies have been extended \cite{Lenzi:2022wjv,Lenzi:2023inn} also to asymptotically de Sitter black holes, either static \cite{Crispino:2013pya} or rotating \cite{CarneirodaCunha:2015qln} and for near-BPS black holes in $d=4, 5$ \cite{Maldacena:1997ih}. Generic fields have been considered in \cite{Boonserm:2008zg}.

In higher dimensions, greybody factors have been computed for asymptotically flat, de Sitter and anti-de Sitter black holes, either static \cite{Harmark:2007jy} or rotating \cite{Jorge:2014kra}. The specific case of graviton emission has been studied in \cite{Cardoso:2005mh}. In brane world scenarios, considering the emitted fields as restricted to live on a 4-dimensional brane, greybody factors have been computed for asymptotically flat \cite{Kanti:2002nr,Harris:2003eg} and de Sitter \cite{Kanti:2014dxa} black holes.

It is of obviously relevance to extend the studies of emission spectra and greybody factors to black holes with higher derivative corrections. In this article, we consider Einstein-Gauss-Bonnet gravity in $d$ dimensions, for which greybody factors have been computed, in the low frequency limit for scalar, fermion and gauge fields for asymptotically flat black holes in \cite{Grain:2005my}. This limit was also taken in the study of scattering problems for the same type of black holes with string corrections in \cite{Moura:2006pz,Moura:2011rr}. These studies have been extended to nonminimally coupled scalar fields and de Sitter $d-$dimensional black holes, also with Gauss-Bonnet corrections, in \cite{Zhang:2017yfu}.

In this article, we will take the opposite (high frequency) limit in two ways: the eikonal (geometrical optics) limit, corresponding to a large real part of the frequency of the emitted radiation; and the asymptotic (highly damped) limit, corresponding to a large imaginary part of such frequency. In a previous article \cite{Moura:2024vhz}, taking the same limits, we have studied greybody factors corresponding to $d-$dimensional black holes with leading string-theoretical $\a$ corrections. The eikonal limit (including greybody factors) has been studied recently for $d=4$ spherically symmetric black holes in Proca-Gauss-Bonnet gravity \cite{Lutfuoglu:2025ldc,Lutfuoglu:2026uzy}.

The article is organized as follows. In section~\ref{sstp}, we review a $d$-dimensional spherically symmetric black hole solution in Einstein-Gauss-Bonnet gravity and its large black hole limit. In Section~\ref{ssbhgp}, we review gravitational perturbations of these black holes, as well as test scalar fields on these backgrounds, writing down the master equations and the corresponding potentials. In Section~\ref{greybody}, we review how the greybody factor is obtained by solving a scattering problem. In section~\ref{eikonal}, we compute the greybody factors corresponding to these perturbations and fields in the eikonal limit. In section~\ref{asymptotic}, we perform an analogous calculation in the highly damped limit using the monodromy method. We conclude by discussing our results in section~\ref{sec:conclusion}.

\section{Einstein-Gauss-Bonnet black holes}
\label{sstp}
\noindent

In this section, we review the $d$-dimensional spherically symmetric black hole solution in Einstein-Gauss-Bonnet gravity and its large black hole limit.

\subsection{The Einstein-Gauss-Bonnet action}
\noindent

We consider the following action defined in $d$ dimensions ($d>4$) corresponding to Gauss-Bonnet gravity:
\be \label{eef} \frac{1}{16 \pi G_d} \int_\mathcal{M} \sqrt{-g} \left( \R + \alpha (\R_{mnpq}\R^{mnpq} - 4\R_{mn}\R^{mn}+ \R^2) \right) \mbox{d}^dx,
\ee
where $G_d$ is the generalized gravitational constant and $\alpha$ is a positive coupling constant. In $d=4$ the Gauss-Bonnet term is topological and does not contribute to the field equations, as is well known.

In the context of superstring theories this action, coupled to a dilaton field, represents the leading correction to Einstein gravity. From this point of view, \eqref{eef} (or its equivalent in string theory) is an effective action, with the parameter $\alpha$ (up to a numerical constant) representing the inverse string tension $\a$.

In Gauss-Bonnet gravity one takes a different point of view, considering~\eqref{eef} as a complete (and not effective) action and computing exact solutions to its field equations (and not just solutions that are perturbative in $\alpha$; in particular, not assuming that $\alpha$ is necessarily a small parameter). Also, in Gauss-Bonnet gravity the dilaton is absent.

\subsection{A spherically symmetric black hole solution}
\noindent

A general static spherically symmetric metric in $d\geq4$ dimensions can always be written depending only on a metric function $f(r)$:
\be \label{schwarz}
\mbox{d}s^2 = -f(r)\ \mbox{d}t^2  + f^{-1}(r)\ \mbox{d}r^2 + r^2 \mbox{d}\Omega^2_{d-2},
\ee
with
\begin{equation}
    \mbox{d}\Omega^2_{d-2}=\sum_{i=1}^{d-2} \prod_{j=1}^{i-1} \sin^2 \theta_j \mbox{d}\theta_i^2.
\end{equation}

In references~\cite{Boulware:1985wk,Wiltshire:1988uq,Wheeler:1985nh,Wheeler:1985qd}, black hole solutions of the field equations from~\eqref{eef} are found and discussed. In this article, we consider a spherically symmetric solution of the form~\eqref{schwarz}, with
\begin{equation} \label{bw}
    f(r) = 1 + \frac{r^2}{\alpha(d-3)(d-4)}(1-q(r)) \hspace{10pt}, \hspace{10pt}q(r) = \sqrt{1 + \frac{2 \alpha (d-3) (d-4) R_0^{d-3}}{r^{d-1}}}.
\end{equation}

The parameter $R_0$ is related to the black hole mass through
\be
M= \frac{(d-2) \Omega_{d-2}}{16 \pi G_d} R_0^{d-3}, \, \Omega_{d-2}=\frac{2 \pi^{\frac{d-1}{2}}}{\Gamma\left(\frac{d-1}{2}\right)}. \label{r0}
\ee
The constant $R_0$ would correspond to the horizon radius of the Tangherlini solution, in the absence of Gauss-Bonnet corrections (i.e. setting $\alpha=0$). The true horizon radius $R_h$ of the black hole~\eqref{bw} is related to $R_0$ through
\be
R_0^{d-3} = R_h^{d-3} \left(1 + \alpha \frac{(d-3)(d-4)}{2 R_h^2}\right). \label{mu}
\ee

\subsection{The perturbative large black hole limit}
\label{plbhl}
\noindent

If the coupling constant $\alpha$ in~\eqref{eef} is small, it is valid to consider it as a perturbative parameter, similarly to $\a$ in string theory. 

More concretely, concerning the black hole solution~\eqref{bw} in which we are interested, defining the dimensionless parameter
\begin{equation}
\lambda' = \frac{\alpha}{R_0^2}, \label{lambda}
\end{equation}
the perturbative limit corresponds to the condition $\lambda' \ll 1$ and, from~\eqref{mu}, it is equivalent to having $\alpha \ll R_h^2$ i.e. the limit of large black holes. We take the perturbative expansion in $f$ and consider only terms to first order in
$\lambda'$ as
\bea
&&f(r) = f_0(r)(1 + \lambda' \delta f(r)), \label{fpert} \\
&&f_0(r) = 1 - \left(\frac{R_0}{r}\right)^{d-3}, \hspace{10pt}  \delta f (r) = \frac{(d-3)(d-4)}{2}
      \left(\frac{R_0}{r}\right)^{2d-4} \frac{1}{1 - \left(\frac{R_0}{r}\right)^{d-3}}. \label{deltf}
\eea

In the large black hole limit, the physical quantities associated with the black hole~\eqref{bw} can also be computed perturbatively in $\lambda'$. The horizon radius $R_h$ reads
\begin{equation}
R_h = R_0 \left(1 -\frac{d-4}{2} \lambda'\right). \label{rh}
\end{equation}
The temperature of a spherically symmetric black hole of the form~\eqref{schwarz}, like~\eqref{bw}, is given by $T_\text{H} = \frac{f'(R_h)}{4\pi}$. In the large black hole limit we are considering, to first order in $\lambda'$, this temperature reads
\begin{equation}
    T_\text{H} = \frac{d-3}{4\pi R_0}\left(1- \lambda' \frac{(d-4)(d-2)}{2}\right). \label{temp}
\end{equation}
%


\section{Gravitational perturbations in the perturbative limit}
\label{ssbhgp}
\noindent

General perturbations $h_{\mu\nu}=\delta g_{\mu\nu}$ of a $d$-dimensional spherically symmetric metric like~\eqref{schwarz} can be uniquely decomposed in their scalar, vectorial and (for $d>4$) tensorial components. Each type of gravitational perturbation is described in terms of master variables $\Psi_a(r,t)$ (the subscript $a$ indicates the kind of perturbation - respectively $\textsf{S}, \textsf{V}, \textsf{T}$). In Einstein gravity, each master variable obeys a second order differential equation (``master equation''):

\be
\frac{\partial^2 \Psi_a}{\partial x^2}(x,t) - \frac{\partial^2 \Psi_a }{\partial t^2}(x,t) = V_a \left[ f(r)\right] \Psi_a(x,t).
\label{pot}
\ee
This ``master equation'' is given in terms of the tortoise coordinate $x$ for the metric~\eqref{schwarz}, defined by
\be
\mbox{d}x = \frac{\mbox{d}r}{f(r)}, \label{tort}
\ee
and of a potential $V_a \left[ f(r) \right]$ that depends on the kind of perturbation one considers~\cite{ik03a}: $V_{\textsf{S}}, V_{\textsf{V}}, V_{\textsf{T}}$. We assume the time dependence of the master variables to be of the form
\be
\Psi_a(x,t) = e^{i\omega t} \psi_a(x), \label{time}
\ee
such that $\frac{\partial\Psi_a}{\partial t} = i\omega \Psi_a$. In this way the master equation~\eqref{pot} may be written in Schr\"odinger form as
\be
\frac{\mbox{d}^2 \psi_a}{\mbox{d} x^2}(x) + \omega^2 \psi_a(x) = V_a \left[ f(r) \right] \psi_a(x).
\label{potential0}
\ee

For static black holes in $d$-dimensional Lovelock gravity of arbitrary order, by perturbing the field equations one also obtains for each perturbation variable a second order master equation like~\eqref{potential0}~\cite{Takahashi:2009xh,Takahashi:2010ye}. Concretely, $d$-dimensional black holes with Gauss-Bonnet corrections have been studied and the potentials have been obtained for tensorial~\cite{Dotti:2005sq,Moura:2012fq},
vectorial and scalar~\cite{Gleiser:2005ra} perturbations. Full expressions for the potentials can be found in~\cite{Moura:2022gqm}, together with the simpler case of a massless scalar test field, which obeys the field equation~\eqref{potential0} with a potential given
by~\cite{Harmark:2007jy}

\be
V_{\textsf{M}} [f(r)] = f(r) \left( \frac{\ell \left( \ell + d - 3 \right)}{r^2} + \frac{\left( d - 2 \right) \left( d - 4 \right) f(r)}{4r^2} + \frac{\left( d - 2 \right) f'(r)}{2r} \right), \label{v0}
\ee
where $\ell$ is the multipole number associated with the spherical harmonic decomposition of the scalar test field. These are the cases we will address in this work in order to compute the greybody factors.

\section{Greybody factors}
\label{greybody}
\noindent

The tortoise coordinate $x$ is defined in~\eqref{tort} in such a way that, for asymptotically flat black holes, $r\to + \infty$ corresponds to $x\to + \infty$ and $r\to R_h^+$ corresponds to $x\to - \infty$. In these regions, the potential $V_a \left[ f(r) \right]$ in~\eqref{potential0} should go to zero:
$$\lim_{r\to + \infty}V_a \left[ f(r) \right] = 0, \, \lim_{r\to + R_h}V_a \left[ f(r) \right] = 0.$$
With these conditions, asymptotically in the regions $x\to - \infty$ and $x \to+ \infty$, the solution $\psi_a(x)$ of the master equation~\eqref{potential0} should have an oscillatory behavior. In order to determine the greybody factors, we seek solutions $\psi_a(x)$ with frequency $\omega$ that obey the boundary conditions
\begin{align}
    \psi_a(x) &\sim T_a(\omega) e^{i\omega x}, \, x\to - \infty, \label{129} \\
\psi_a(x) &\sim e^{i\omega x}+ R_a(\omega)e^{-i\omega x}, \, x\to + \infty, \label{130}
\end{align}
where the complex numbers $T_a(\omega), R_a(\omega)$ are called transmission and reflection coefficients, respectively.

Physically, $\psi_a(x)$ describes the scattering of an incoming wave, originating at spatial infinity, off the non-trivial structure of the potential $V_a[f(r)]$, which results from the curvature of spacetime. In general, a scattering problem with a potential allows for solutions with complex frequency $\omega$. In this case, the real part of $\omega$ gives the oscillation frequency of the wave, while the imaginary part governs the growth or decay of its amplitude in time. In order to have damped solutions one should have $\Im\left(\omega\right)>0$ (recall equation~\eqref{time}).


If $\psi_a(x)$ is a solution for some fixed $\omega$, then the function $\widetilde{\psi}_a(x)$ obtained from $\psi_a(x)$ by replacing $\omega \to -\omega$ also solves~\eqref{potential0}, since this equation is invariant under $\omega \mapsto -\omega$. It does, however, satisfy different boundary conditions, namely
\begin{align}
    \widetilde{\psi}_a(x) &\sim \widetilde{T}_a(\omega) e^{-i \omega x}, \, x \to - \infty, \label{130til} \\
\widetilde{\psi}_a(x) &\sim e^{-i \omega x} + \widetilde{R}_a(\omega) e^{i\omega x},\, x \to + \infty, \label{129til}
\end{align}
for some other reflection and transmission coefficients $\widetilde{R}_a(\omega) := R_a(-\omega)$, $\widetilde{T}_a(\omega) := T_a(-\omega).$ One can easily show that the flux (or Wronskian)
\begin{equation}
    J_a(x) = \frac{1}{2i} \left( \widetilde{\psi}_a(x)\frac{\mbox{d} \psi_a}{\mbox{d}x}(x) - \psi_a(x)\frac{\mbox{d} \widetilde{\psi}_a}{\mbox{d}x}(x) \right)
\end{equation}
does not depend on $x$, i.e. $\frac{\text{d} J_a}{\text{d}x}=0$. Evaluating such flux at both $x \to \pm \infty$ yields the condition (valid for all asymptotically flat spacetimes~\cite{Harmark:2007jy})
\begin{equation}
R_a(\omega) \widetilde{R}_a(\omega) + T_a(\omega) \widetilde{T}_a(\omega) = 1. \label{148}
\end{equation}
The greybody factor is defined as
\begin{equation}
    \gamma_a(\omega) = T_a(\omega)\widetilde{T}_a(\omega).
    \label{149}
\end{equation}

If the frequency $\omega$ is real, $\widetilde{\psi}_{a}(x)=\psi_a(x)^*$, $\widetilde{R}_a(\omega)= R_a(\omega)^\ast$ and $\widetilde{T}_a(\omega) =T_a(\omega)^\ast$, in which case one has the familiar scattering formula
\begin{equation}
\left|R_a(\omega)\right|^2 + \left|T_a(\omega)\right|^2 = 1, \, \,
\gamma_a(\omega) = \left|T_a(\omega)\right|^2.
\label{gfd}
\end{equation}
Integrating Hawking's formula~\eqref{gfdef} over the whole frequency spectrum, and interpreting the result as a radiation emission rate, only makes sense for real frequencies $\omega$ of the emitted radiation. In this work, however, we take the emitted radiation to have complex frequency, in which case the interpretation of~\eqref{gfdef}, beyond that of an analytic continuation, is less clear. An exception is the limit of purely imaginary $\omega$: since the imaginary part of $\omega$ describes radiation damping, the integral of~\eqref{gfdef} over the whole frequency spectrum can then be interpreted as a radiation decay rate.

In this article, we compute analytical expressions for the greybody factors associated with gravitational perturbations and test scalar fields for $d$-dimensional black holes with Gauss-Bonnet higher derivative corrections given by~\eqref{bw}, in the perturbative large black hole limit described in subsection~\ref{plbhl}. We consider two different limits: the eikonal limit of large multipole number $\ell$ and the asymptotic limit in which we restrict our attention to frequencies whose imaginary part is large in magnitude and much larger than the real part, that is, almost purely imaginary frequencies.

\section{Greybody factors in the eikonal limit}
\label{eikonal}
\noindent

In the eikonal limit one can use the WKB method in order to solve scattering problems described by equations like~\eqref{potential0}, as long as the associated potential $V_a\left[f(r)\right]$ has one single peak (maximum). That is the case for the potentials corresponding to test scalar fields and gravitational perturbations in Einstein gravity, and also in the presence of Gauss-Bonnet corrections. Indeed, as we consider $\lambda'$ to be a small perturbative parameter, the higher derivative corrections we consider do not change the shape of these potentials. In this limit, the quasinormal modes corresponding to the black hole solution we have been considering were obtained in \cite{Konoplya:2017wot}. Quasinormal modes corresponding to an analogous $d-$dimensional black hole solution with leading string-theoretical higher derivative corrections, derived by Callan, Myers and Perry \cite{cmp89}, have been obtained in \cite{Moura:2021eln}.

Greybody factors in the eikonal limit, for spherically symmetric $d$-dimensional black holes in Einstein gravity, have been obtained in~\cite{Konoplya:2019hlu,Konoplya:2023moy}. The method relies on a function $U_a(x,\omega)=V_a(x)-\omega^2$ and its derivatives, $V_a(x)$ being the potential $V_a\left[f(r)\right]$ in~\eqref{potential0}, but given in terms of the tortoise coordinate $x$. If $x_0$ is the point where the potential $V_a(x)$ reaches a maximum, we define
\be
U_{a,0}(\omega) = U_a(x_0,\omega), \, U_{a,2}(\omega) = \left.\frac{\mbox{d}^2 U_a(x,\omega)}{\mbox{d} x^2}\right|_{x=x_0}. \label{uw}
\ee
The greybody factor is then simply given by
\be
\gamma_a(\omega)=\frac{1}{1+e^{2\pi i \kappa_a}}, \,\, \kappa_a=-i\frac{U_{a,0}(\omega)}{\sqrt{-2 U_{a,2}(\omega)}}. \label{gkk}
\ee
According to our perturbative approach, $\gamma_a(\omega)$ can be expanded in $\lambda'$ as
\begin{equation}
    \gamma_a(\omega) = \gamma_{0}(\omega)\left(1+\lambda' \delta\gamma_a(\omega)\right). \label{geik}
\end{equation}
In this expression, $\gamma_0(\omega)$ is the eikonal greybody factor in Einstein gravity, without $\lambda'$ corrections. It depends only on the eikonal limit of the uncorrected potential $V_{\textsf{M}}[f(r)]$, given by~\eqref{v0}, since this limit is the same for all gravitational perturbations and for scalar test fields~\cite{ik03a}. The result for $\gamma_0(\omega)$ is therefore universal, and given by
\be
\gamma_0(\omega) = \left(\exp \left(\frac{\pi \ell}{\sqrt{d-3}}-\frac{4^{\frac{1}{3-d}-2} \sqrt{d-3} (d-1)^{\frac{2}{d-3}+1}}{\pi \ell} \frac{\omega^2}{T_\text{H}^2} \right)+1\right)^{-1}. \label{geik0}
\ee
In the eikonal (large $\ell$) limit, besides the $\lambda'$ expansion one can also consider an expansion in $1/\ell$. We have then $1/\ell$ corrections to the $\lambda'$ correction $\delta \gamma_a(\omega)$ in (\ref{geik}), which can be written as
\begin{equation}
    \delta \gamma_a(\omega) = \Gamma(\omega)\left(\Gamma_{a,0}(\omega)\ell +\Gamma_{a,1}(\omega)\frac{1}{\ell} + \mathcal{O}\left(\frac{1}{\ell^2}\right) \right). \label{dgeik}
\end{equation}
The term $\Gamma(\omega)$ is also the same for gravitational perturbations and scalar test fields, and given by
\be
\Gamma(\omega) = \left(\exp \left(\frac{4^{\frac{1}{3-d}-2} \sqrt{d-3} (d-1)^{\frac{2}{d-3}+1}}{\pi \ell} \frac{\omega^2}{T_\text{H}^2}-\frac{\pi  \ell}{\sqrt{d-3}}\right)+1\right)^{-1}. \label{geik1}
\ee

We now compute $\Gamma_{a,0}(\omega)$ and $\Gamma_{a,1}(\omega)$ separately for each $a  = \textsf{S}, \textsf{V}, \textsf{T}, \textsf{M}$. For tensor-type gravitational perturbations, we can write
\begin{equation}
    U_{\textsf{T},0}(\omega) = \frac{2^{\frac{1}{d-3}} (d-3) (d-1)^{-\frac{4}{d-3}-2}\ell^2 \left(2^{\frac{1}{d-3}} (d-1)^{\frac{2}{d-3}+1} (1-(d-4) \lambda' )+3\ 2^{\frac{d}{d-3}} (d-2) \lambda' \right)}{R_h^2}-\omega ^2
\end{equation}
and
\begin{align}
     &U_{\textsf{T},2}(\omega) = \frac{2 (d-3)^3 (d-1)^{-\frac{7}{d-3}-3}\ell^2}{R_h^4}\times \\& \hspace{3cm} \times\left\{2^{\frac{d}{d-3}} \left(8^{\frac{1}{d-3}} (d-9) d+7\ 2^{\frac{d}{d-3}}\right) (d-1)^{\frac{1}{d-3}} \lambda' +16^{\frac{1}{d-3}} (d-1)^{\frac{d}{d-3}} (2 (d-4) \lambda' -1)\right\}, \nonumber
\end{align}
from which we get
\begin{equation}
   \Gamma_{\textsf{T},0}(\omega) = -\frac{\pi  2^{\frac{2}{d-3}} (d-2) (d-1)^{-\frac{2}{d-3}}}{\sqrt{d-3}}
\end{equation}
and
\begin{align}
      \Gamma_{\textsf{T},1}(\omega) = \frac{2^{-\frac{8}{d-3}-4} (d-1)^{-\frac{2}{d-3}}}{\pi  \sqrt{d-3}  T_{\text{H}}^2}\Bigg\{(d-1)^{\frac{d+1}{d-3}} \left(64^{\frac{1}{d-3}} (d-7) d+3\times 4^{\frac{d}{d-3}}\right) \omega ^2 +(d-1)^{\frac{2}{d-3}} \Big(-3\times 4^{\frac{d+1}{d-3}} d^2\nonumber\\+256^{\frac{1}{d-3}} d \left(d^2+41\right)-\left(\left(64^{\frac{1}{d-3}} d (d ((d-9) d+27)-31)+3\times 4^{\frac{d}{d-3}}\right) (d-1)^{\frac{2}{d-3}}\right)-21\ 2^{\frac{d+5}{d-3}}\Big) \omega ^2\Bigg\}.
\end{align}
For gravitational vector type perturbations, we can write
\begin{equation}
    U_{\textsf{V},0}(\omega) = \frac{2^{\frac{1}{d-3}} (d-3) (d-1)^{-\frac{4}{d-3}-2}\ell^2 \left(2^{\frac{1}{d-3}} (d-1)^{\frac{2}{d-3}+1} (1-(d-4) \lambda' )-2^{\frac{d}{d-3}} (d-4) (d-2) \lambda' \right)}{R_h^2}-\omega ^2
\end{equation}
and
\begin{align}
    &U_{\textsf{V},2}(\omega) = \frac{2^{\frac{d}{d-3}} (d-3)^3 (d-1)^{-\frac{7}{d-3}-3}\ell^2}{R_h^4} \times \\ &\hspace{4cm}\times\left(2^{\frac{d}{d-3}+1} (d-4) (d-2) (d-1)^{\frac{1}{d-3}} \lambda' +2^{\frac{1}{d-3}} (d-1)^{\frac{d}{d-3}} (2 (d-4) \lambda' -1)\right), \nonumber
\end{align}
from which we get
\begin{equation}
    \Gamma_{\textsf{V},0}(\omega) = 0
\end{equation}
and
\begin{equation}
\begin{split}
    \Gamma_{\textsf{V},1}(\omega) = \frac{\left(2-4^{\frac{1}{3-d}} (d-1)^{\frac{2}{d-3}+1}\right) (d-4) \sqrt{d-3} (d-2) \omega ^2}{16 \pi   T_\text{H}^2}.
\end{split}
\end{equation}
For gravitational scalar type perturbations, we can write
\begin{equation}
    U_{\textsf{S},0}(\omega) = \frac{2^{\frac{1}{d-3}} (d-3) (d-1)^{-\frac{4}{d-3}-2}\ell^2 \left(2^{\frac{1}{d-3}} (d-1)^{\frac{2}{d-3}+1} (1-(d-4) \lambda' )-2^{\frac{d}{d-3}} (d-4) (2 d-3) \lambda' \right)}{R_h^2}-\omega ^2
\end{equation}
and
\begin{align}
    & U_{\textsf{S},2}(\omega) =  \frac{2^{\frac{1}{d-3}} (d-3)^3 (d-1)^{-\frac{7}{d-3}-3}\ell^2}{R_h^4}\times \\ &\hspace{3cm
    }\times \left(2^{\frac{5}{d-3}+2} (d-4) (3 d-5) (d-1)^{\frac{1}{d-3}} \lambda' +(d-1)^{\frac{d}{d-3}} \left(2^{\frac{3}{d-3}+2} (d-4) \lambda' -2^{\frac{d}{d-3}}\right)\right), \nonumber
\end{align}
from which we get
\begin{equation}
    \Gamma_{\textsf{S},0}(\omega) =\frac{\pi  2^{\frac{2}{d-3}} (d-4) (d-1)^{-\frac{2}{d-3}} }{\sqrt{d-3}}
\end{equation}
and
\begin{align}
   & \Gamma_{\textsf{S},1}(\omega) =\frac{2^{-\frac{5}{d-3}-4} (d-4)}{\pi  \sqrt{d-3}  T_\text{H}^2} \Big\{3\times 32^{\frac{1}{d-3}} \left(d^2+5\right)+\\& \hspace{4cm}+\left(2^{\frac{d}{d-3}} ((d-3) d+3)-8^{\frac{1}{d-3}} d ((d-4) d+5)\right) (d-1)^{\frac{2}{d-3}}-7\ 2^{\frac{d+2}{d-3}} d\Big\} \omega ^2.\nonumber
\end{align}
Finally, for scalar test fields, we may write
\begin{equation}
    U_{\textsf{M},0}(\omega) = \frac{2^{\frac{1}{d-3}} (d-3) (d-1)^{-\frac{4}{d-3}-2}\ell^2 \left(2^{\frac{1}{d-3}} (d-1)^{\frac{2}{d-3}+1} (1-(d-4) \lambda' )+2^{\frac{d}{d-3}} (d-4) \lambda' \right)}{R_h^2}-\omega ^2
\end{equation}
and
\begin{align}
    &U_{\textsf{M},2}(\omega) = \frac{2^{\frac{1}{d-3}} (d-3)^3 (d-1)^{-\frac{6}{d-3}-3}\ell^2}{R_h^4}\times\\ &\hspace{4.5cm}\times \left((d-1)^{\frac{2}{d-3}+1} \left(2^{\frac{3}{d-3}+2} (d-4) \lambda' -2^{\frac{d}{d-3}}\right)+2^{\frac{5}{d-3}+2} (d-4) (d-3) \lambda' \right), \nonumber
\end{align}
from which we get
\begin{equation}
   \Gamma_{\textsf{M},0}(\omega) =-\frac{\pi  2^{\frac{2}{d-3}} (d-4) (d-1)^{-\frac{2}{d-3}}}{\sqrt{d-3}}
\end{equation}
and
\begin{equation}
\begin{split}
   \Gamma_{\textsf{M},1}(\omega) =\frac{2^{-\frac{5}{d-3}-4} (d-4)}{\pi  \sqrt{d-3}  T_\text{H}^2} \left(\left(3\times 2^{\frac{d}{d-3}} \left(d^2+1\right)-8^{\frac{1}{d-3}} d \left(d^2+11\right)\right) (d-1)^{\frac{2}{d-3}}+2^{\frac{5}{d-3}} (d-3)^2\right) \omega ^2.
\end{split}
\end{equation}


\section{Greybody factors in the highly damped limit}

\label{asymptotic}

\noindent


In this section, we compute the greybody factors in the highly damped limit, assuming we are working with frequency values $\omega$ whose imaginary part is large and much larger than the real part (almost purely imaginary). We will use the monodromy method developed in~\cite{Motl:2003cd,Neitzke:2003mz,Keshet:2007be,Gaiotto:2012rg}. In the perturbative large black hole limit, the $d$-dimensional Gauss--Bonnet black hole reduces to the spacetime defined by~\eqref{fpert} and~\eqref{deltf}, which is structurally analogous to the Callan--Myers--Perry black hole~\cite{cmp89}. The monodromy method was already used in~\cite{Moura:2021nuh} to compute the quasinormal modes corresponding to the Callan--Myers--Perry solution and in \cite{Moura:2022gqm} to compute the quasinormal modes corresponding to the $d-$dimensional Gauss-Bonnet black hole solution we have been considering.

This method was also used in~\cite{Moura:2024vhz} to compute the greybody factors in the highly damped limit for the Callan--Myers--Perry solution and, owing to this structural similarity, its implementation there carries over directly to the present case. We therefore only outline the method, its key steps and the final results, and refer the reader to~\cite{Moura:2024vhz} for further details.

The strategy we use to compute the greybody factor consists in building a linear system of three algebraic equations where one of the independent variables is the reflection coefficient $R_a(\omega)$ (where once again $a=\textsf{S},\textsf{V},\textsf{T},\textsf{M}$). After solving this system and consequently finding $R_a(\omega)$, we repeat the process to find $\widetilde{R}_a(\omega)= R_a(-\omega)$. Finally, we relate these coefficients with the greybody factor $\gamma_a(\omega)$, using the equation
\begin{equation}
   R_a(\omega)\widetilde{R}_a(\omega) + \gamma_a(\omega) = 1,
   \label{30}
\end{equation}
which follows directly from the flux conservation~\eqref{148} and equation~\eqref{149}.

\subsection{Computation of $R_a(\omega)$}
\noindent

The linear system we have to build in order to solve for the reflection coefficient $R_a(\omega)$ consists of three equations, each obtained as a consistency condition obeyed by the solution $\psi_a(x)$ of the master equation~\eqref{potential0}. The first step consists in performing a small $\lambda'$ perturbative expansion of the corresponding potentials and solution:
\begin{align}
V_a(x) &= V_a^0(x) + \lambda' V_a^1(x),   \\
    \psi_a(x) &= \psi_{a,0}(x) + \lambda' \psi_{a,1}(x), \hspace{3pt} a=\textsf{S},\textsf{V},\textsf{T},\textsf{M}.
\end{align}
Using these expansions, the master equation~\eqref{potential0} splits into two equations: a zeroth-order homogeneous equation, whose solution is $\psi_{a,0}(x)$, and a first-order non-homogeneous equation, whose solution is $\psi_{a,1}(x)$. These correspond to equations~(4.13) and~(4.14) of~\cite{Moura:2022gqm}, where the respective potentials and non-homogeneous terms are derived in section~3 therein.

Near $r = 0$, the zeroth order equation reads
\begin{equation}
    \frac{\mbox{d}^2 \psi_{a,0}}{\mbox{d} z^2}(z) + \left(\omega^2 -\frac{j^2-1}{4z^2}\right)\psi_{a,0}(z) = 0,
\end{equation}
where $j = 0$ for $a=\textsf{S},\textsf{T},\textsf{M}$ and $j = 2$ for $a=\textsf{V}$. The variable $z$ corresponds to the tortoise coordinate of the Tangherlini solution, defined as (recall equation~\eqref{deltf})
\begin{equation}
    \mbox{d} z = \frac{\mbox{d}r}{f_0(r)},
\end{equation}
for an integration constant chosen so as to ensure that $x$ converges to $z$ for large $|r|$. The generic solution can be written as
\begin{equation}
    \psi_{a,0}(z) = A_+ \sqrt{\omega z 2\pi}\text{J}_{\frac{j}{2}}\left(\omega z\right) + A_- \sqrt{\omega z 2\pi}\text{J}_{-\frac{j}{2}}\left(\omega z\right), \hspace{3pt} A_+,A- \in \mathbb{C}.
    \label{eq:nearoriginsolution}
\end{equation}

\begin{figure}[h]
    \centering
    \includegraphics[width=0.4\linewidth]{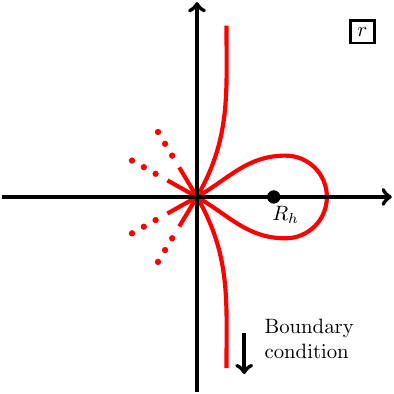}
    \caption{Schematic representation of the Stokes lines associated with the master equation~\eqref{potential0}, expressed in terms of the tortoise coordinate $z$. Stokes lines are shown as solid red lines; truncated lines ending in dots indicate additional ($d$-dependent) Stokes lines that are not relevant to our analysis. The asymptotic direction along which we enforce the boundary condition~\eqref{130} is indicated by a black arrow.}
    \label{fig:stokes}
\end{figure}

In order to enforce the defining greybody factor boundary data~\eqref{129} and~\eqref{130}, we need to understand how the solution above behaves globally in $r$, particularly about the event horizon ($x \to -\infty$) and asymptotic infinity ($x \to +\infty$). As argued for in~\cite{Motl:2003cd}, both tasks are dramatically simplified when we frame our problem in the complex $r$-plane:

\begin{itemize}

\item First, working in the complex $r$-plane allows us to use WKB methods to track the solution~\eqref{eq:nearoriginsolution} analytically away from $r = 0$. More concretely, the WKB approximation can be applied reliably along Stokes lines\footnote{Throughout, Stokes lines denote the curves along which $\omega z$ is real, so that the WKB exponentials in~\eqref{130} are purely oscillatory (recall that $x$ converges to $z$ for large $|r|$). In the mathematical literature these are usually called anti-Stokes lines.}, which extend globally in the complex $r$-plane and generically take the shape depicted in figure~\ref{fig:stokes} for all $a=\textsf{S},\textsf{V},\textsf{T},\textsf{M}$ and $d > 4$.

    \item Second, the boundary conditions at asymptotic infinity may equivalently be imposed along a Stokes line of the spectral network~\cite{Gaiotto:2012rg} associated with the master equation~\eqref{potential0} that extends to infinity. Along such lines, both exponential terms in~\eqref{130} are purely oscillatory and of comparable magnitude, so neither dominates the other and both can be reliably tracked within the WKB approximation. Following~\cite{Moura:2024vhz}, we choose to impose the boundary condition~\eqref{130} along the lowermost unbounded Stokes line as $|r| \to +\infty$, shown in figure~\ref{fig:stokes}.

    \item Finally, the boundary condition~\eqref{129} can be implemented by imposing the corresponding monodromy of $\psi_a(x)$ around the event horizon $r = R_h$.
\end{itemize}

Analytically continuing the solution~\eqref{eq:nearoriginsolution} along the lowermost unbounded Stokes line depicted in figure~\ref{fig:stokes}, results in (see~\cite{Moura:2021nuh} for further details)
\begin{align}
     &\psi_a(z)  \sim
    \left(A_+e^{i\alpha_+}+A_-e^{i\alpha_-}\right)e^{-i\omega z}\left[1 + \lambda' \left(\frac{\left(\Lambda_I^+\right)_a e^{i\alpha_+} +\left(\Lambda_I^-\right)_ae^{i\alpha_-}}{A_+e^{i\alpha_+} +A_-e^{i\alpha_-}}\right)\right]+ \label{173}\\ &\hspace{5cm}+
    \left(A_+e^{-i\alpha_+}+A_-e^{-i\alpha_-}\right)e^{i\omega z}\left[1 + \lambda' \left(\frac{\left(\Lambda_I^+\right)_a e^{-i\alpha_+} +\left(\Lambda_I^-\right)_ae^{-i\alpha_-}}{A_+e^{-i\alpha_+} +A_-e^{-i\alpha_-}}\right)\right],
    \nonumber
\end{align}
for large $|r|$ where
\begin{equation}
    \alpha_ {\pm} \coloneqq \frac{\pi}{4}\left(1 \pm j\right) \label{105}.
\end{equation}
The constants $(\Lambda_I^\pm)_a$ coincide with the constants $\Lambda_I^\pm$ defined in equation~(3.55) of~\cite{Moura:2021nuh}, upon the replacement
\begin{equation}
    \xi_3 \to (\xi_a)_3\,,
    \label{23}
\end{equation}
where the coefficients $\xi_a$, with $a = \textsf{S}, \textsf{V}, \textsf{T}, \textsf{M}$, are given in equation~(4.18) of~\cite{Moura:2022gqm}. Enforcing the boundary condition~\eqref{130} on~\eqref{173} yields
\begin{align}
    R_a(\omega) &= \left(A_++\lambda'(\Lambda_I^+)_a\right)e^{+i\alpha_+} + \left(A_-+\lambda'(\Lambda_I^-)_a\right)e^{+i\alpha_-},
   \label{24} \\  1 &=\left(A_++\lambda'(\Lambda_I^+)_a\right)e^{-i\alpha_+} + \left(A_-+\lambda'(\Lambda_I^-)_a\right)e^{-i\alpha_-}.
   \label{25}
\end{align}
The equations above make up two of the three algebraic equations of the system we wish to solve.

\begin{figure}[h]
    \centering
\includegraphics[width=0.55\linewidth]{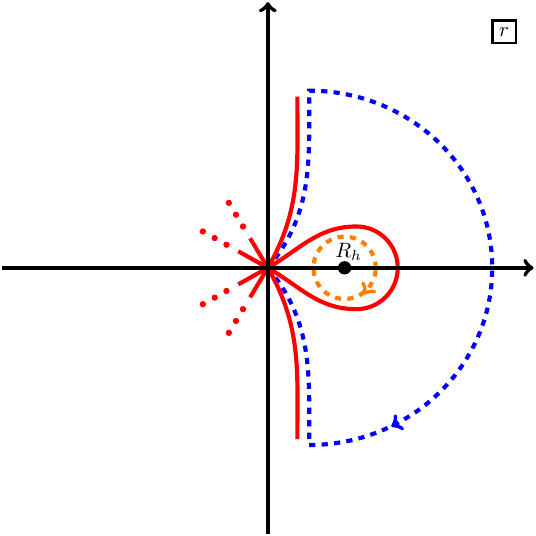}
    \caption{Schematic representation of the closed contours (dashed blue and orange lines) along which $\psi_a(x)$ is analytically continued in order to impose the monodromy matching condition of~\cite{Motl:2003cd}. Stokes lines are shown as solid red lines; truncated lines ending in dots indicate additional ($d$-dependent) Stokes lines that are not relevant to our analysis.}
    \label{fig:contours}
\end{figure}

Following~\cite{Moura:2024vhz}, the last one can be obtained by the monodromy matching condition~\cite{Motl:2003cd}. More concretely, we start by analytically continuing the solution~\eqref{173} along the blue contour depicted in figure~\ref{fig:contours}. Doing so results in the multiplicative monodromy term (see~\cite{Moura:2022gqm} for further details)
\begin{equation}
    \mathcal{N}_a := \left(\frac{A_+e^{5i\alpha_+} + A_-e^{5i\alpha_-}}{A_+e^{i\alpha_+} + A_-e^{i\alpha_-}}\right) e^{-i\omega \Delta_z}\left(1 + \lambda' \delta \mathcal{N}_a\right).
    \label{137}
\end{equation}
The constant $\delta \mathcal{N}_a$ reads
\begin{equation}
    \delta \mathcal{N}_a := \frac{\left(\Lambda_G^+\right)_ae^{5i\alpha_+} +\left(\Lambda_G^-\right)_ae^{5i\alpha_-}}{A_+ e^{5i\alpha_+}+ A_-e^{5i\alpha_-}} - \frac{\left(\Lambda_I^+\right)_ae^{i\alpha_+} +\left(\Lambda_I^-\right)_ae^{i\alpha_-}}{A_+e^{i\alpha_+} + A_-e^{i\alpha_-}},
\end{equation}
where the constants $(\Lambda_G^\pm)_a$ coincide with the constants $\Lambda_G^\pm$ defined in equation~(5.39) of~\cite{Moura:2022gqm}, upon the replacement~\eqref{23}, and
\begin{equation}
    \Delta_z = -2\pi i \frac{R_h}{d-3}.
\end{equation}

The next step consist in computing the multiplicative monodromy of $\psi_a(x)$ around the orange contour depicted in figure~\ref{fig:contours}. This is a simpler task which results in (see~\cite{Moura:2022gqm} for further details)
\begin{equation}
    \mathcal{M}_a =e^{i \omega \Delta_x}, \hspace{3pt} \Delta_x = \Delta_z\left(1+\frac{\lambda'}{2}(d-4)(d-2)\right).
\end{equation}
Since the orange contour can be smoothly deformed into the blue one (see figure~\ref{fig:contours}) without crossing any singularity of the master equation~\eqref{potential0} (namely, the real and fictitious event horizons and $r = 0$), the two monodromies computed above must coincide. Imposing this equality yields the monodromy matching condition
\begin{equation}
   R_a(\omega) = e^{-i\omega(\Delta_z + \Delta_x)}\left[\left(A_+ + \lambda'(\Lambda_G^+)_a\right)e^{5i\alpha_+} + \left(A_- + \lambda'(\Lambda_G^-)_a\right)e^{5i\alpha_-}\right],
   \label{26}
\end{equation}
which, together with~\eqref{24} and~\eqref{25}, completes our linear system.

In order to solve this system, we use standard perturbation theory, considering the expansions
\begin{align}
     R_a(\omega) &= R_{a,0}(\omega) + \lambda'R_{a,1}(\omega), \\
      A_\pm &= A_\pm^0+\lambda'A_\pm^1.
\end{align}
Replacing these expansions in the system above and solving perturbatively in powers of $\lambda'$ yields two algebraic systems. The first one, of zeroth order in $\lambda'$, is
\begin{equation}
    \begin{dcases}  A_+^0e^{5i\alpha_+} + A_-^0e^{5i\alpha_-} =   R_{a,0}(\omega) e^{\frac{\omega}{T_{\text{H}}}}\\
    A_+^0e^{i\alpha_+} + A_-^0e^{i\alpha_-} =  R_{a,0}(\omega) \\
    A_+^0e^{-i\alpha_+} + A_-^0e^{-i\alpha_-} = 1
    \end{dcases},
\end{equation}
whose solution is
\begin{equation}
     R_{a,0}(\omega) =  \frac{2i}{e^{\frac{\omega}{T_{\text{H}}}}+2\cos(\pi j)+1}\cos\left(\frac{\pi j}{2}\right)
\end{equation}
for all $a = \textsf{S}, \textsf{V}, \textsf{T}, \textsf{M}$ and
\begin{align}
     A_+^0 &= \frac{e^{i\alpha_-} - R_{a,0}(\omega)e^{-i\alpha_-}}{2 i \sin\left(\alpha_--\alpha_+\right)}, \\
      A_-^0&= \frac{R_{a,0}(\omega)e^{-i\alpha_+}- e^{i\alpha_+}}{2i\sin\left(\alpha_--\alpha_+\right)}.
\end{align}

The second algebraic system, of first order in $\lambda'$, is
\begin{equation}
    \begin{dcases}  \left(A_+^1 + (\sigma_+)_a\right)e^{5i\alpha_+} + \left(A_-^1+ (\sigma_-)_a\right)e^{5i\alpha_-} =   e^{\frac{\omega}{T_{\text{H}}}} \left(R_{a,1}(\omega) - L \omega T_{\text{H}}^{-1} R_{a,0}(\omega)\right)\\
    \left(A_+^1+ (\zeta_+)_a\right)e^{i\alpha_+} + (A_-^1 + (\zeta_-)_a)e^{i\alpha_-} = R_{a,1}(\omega) \\
    \left(A_+^1+ (\zeta_+)_a\right)e^{-i\alpha_+} + (A_-^1 + (\zeta_-)_a)e^{-i\alpha_-} = 0
    \end{dcases},
\end{equation}
where $(\sigma_\pm)_a$ and $(\zeta_\pm)_a$ are respectively identical to the constants $\sigma_\pm$ and $\zeta_\pm$ defined in equations~(5.57) and~(5.58) of~\cite{Moura:2024vhz}, under the redefinition~\eqref{23}. Furthermore, we defined the constant
\begin{equation}
    L = \frac{(d-4)(d-2)}{4}.
\end{equation}

Solving this system for $R_{a,1}(\omega)$ yields
\begin{equation}
    R_{1,a}(\omega) = -\frac{1}{e^{\frac{\omega}{T_{\text{H}}}}+2\cos(\pi j)+1}\left(2(\Sigma_1)_a\cos(\pi j)-2i(\Sigma_2)_a\cos\left(\frac{\pi j}{2}\right) + (\Sigma_1)_a + (\Sigma_3)_a\right).
\end{equation}
where the constants $(\Sigma_1)_a$, $(\Sigma_2)_a$ and $(\Sigma_3)_a$ are respectively identical to the constants $\Sigma_1$, $\Sigma_2$ and $\Sigma_3$ defined in equations (5.60), (5.61) and (5.62) of~\cite{Moura:2024vhz}, under the redefinition~\eqref{23}).

After some algebraic manipulation, we can rewrite $R_a(\omega)$ as
\begin{equation}
    R_a(\omega) = R_{a,0}(\omega)\left(1 + \lambda'\delta R_a(\omega)\right),
\end{equation}
where
\begin{equation}
    \delta R_a(\omega) = -\frac{1}{R_{a,0}(\omega)\left(e^{\frac{\omega}{T_{\text{H}}}}+2\cos(\pi j)+1\right)}\left(2(\Sigma_1)_a\cos(\pi j)-2i(\Sigma_2)_a\cos\left(\frac{\pi j}{2}\right) + (\Sigma_1)_a + (\Sigma_3)_a\right).
\end{equation}
Taking the limit $j\to 0$ for $a = \textsf{S},\textsf{T},\textsf{M}$ and $j\to 2$ for $a = \textsf{V}$ yields
\begin{align}
     R_{a,0}(\omega) &= \frac{2i}{e^{\frac{\omega}{T_{\text{H}}}}+3}, \hspace{3pt} a = \textsf{S},\textsf{T},\textsf{M}, \\
      R_{\textsf{V},0}(\omega) &=- \frac{2i}{e^{\frac{\omega}{T_{\text{H}}}}+3}.
\end{align}
Furthermore, after some algebraic manipulation, we get
\begin{equation}
    \delta R_a(\omega) =\frac{1}{e^{\frac{\omega}{T_{\text{H}}}}+3} \left[\frac{(d-4)(d-2)}{4}\frac{\omega}{T_{\text{H}}}e^{\frac{\omega}{T_{\text{H}}}}+ \left(\frac{\omega}{T_{\text{H}}}\right)^{\frac{d-1}{d-2}}\left(\frac{d-3}{4\pi}\right)^{\frac{d-1}{d-2}}\Upsilon_a\right],
\end{equation}
where we defined the constants
\begin{align}
  &\Upsilon_{\textsf{M}} = e^{-\frac{3\pi i}{2}\left(1 + \frac{1}{d-2}\right)}\frac{\pi ^2 (2d-4)^{-\frac{1}{d-2}} (d-4) (d-3)   \Gamma \left(\frac{1}{d-2}\right) \left(e^{-\frac{i\pi}{d-2}} \left(e^{\frac{\omega}{T_{\text{H}}}}-1\right)+4\right)}{(d-1) \Gamma \left(\frac{1}{2} + \frac{1}{2(d-2)}\right)^4}, \\
     &\Upsilon_a = e^{-\frac{3 \pi i}{2}\left(1+ \frac{1}{d-2}\right)} \left(e^{-\frac{i \pi}{d-2}} \left(e^{\frac{\omega}{T_{\text{H}}}}-1\right)+4\right)\times \\ & \hspace{5cm}\times \frac{\pi ^2 (2d-4)^{-\frac{1}{d-2}} (d-4) (d-1)^3 ((d-5) d+2)  \Gamma \left(\frac{1}{d-2}\right)}{16 (d-2)^5 \Gamma \left(\frac{1}{2}\left(3 + \frac{1}{d-2}\right)\right)^4}, \hspace{3pt} a = \textsf{S},\textsf{T},\textsf{V}. \nonumber
\end{align}
%

\subsection{Computation of $\widetilde{R}_a(\omega)$}
\noindent

The computation of $\widetilde{R}_a(\omega)$ follows identically to the previous one, upon replacing the boundary conditions~\eqref{129} and~\eqref{130} by~\eqref{130til} and~\eqref{129til}. The first two algebraic equations of the linear system read
\begin{align}
     1 &= \left(\widetilde{A}_+ +\lambda'(\widetilde{\Lambda}_I^+)_a\right)e^{i\alpha_+} + \left(\widetilde{A}_-+\lambda'(\widetilde{\Lambda}_I^-)_a\right)e^{i\alpha_-},
   \label{27} \\
\widetilde{R}_a(\omega) &= \left(\widetilde{A}_++\lambda'(\widetilde{\Lambda}_I^+)_a\right)e^{-i\alpha_+} + \left(\widetilde{A}_-+\lambda'(\widetilde{\Lambda}_I^-)_a\right)e^{-i\alpha_-},
   \label{28}
\end{align}
where the constants $(\widetilde{\Lambda}_I^\pm)_a$ are obtained from $(\Lambda_I^\pm)_a$ by replacing $A_\pm$ with $\widetilde{A}_\pm$. The monodromy matching equation reads
\begin{equation}
   1 = e^{-i\omega( \Delta_z- \Delta_x)}\left[\left(A_+ +\lambda'(\widetilde{\Lambda}_G^+)_a\right)e^{5i\alpha_+} +  \left(A_-+ \lambda'(\widetilde{\Lambda}_G^-)_a\right)e^{5i\alpha_-}\right],
      \label{29}
\end{equation}
where $(\widetilde{\Lambda}_G^\pm)_a$ are again obtained from $(\Lambda_G^\pm)_a$ by replacing $A_\pm$ with $\widetilde{A}_\pm$.

In order to solve the algebraic system composed of equations (\ref{27}), (\ref{28}) and (\ref{29}), we use standard perturbation theory, considering the expansions
\begin{align}
      \widetilde{R}_a(\omega) &= \widetilde{R}_{a,0}(\omega) + \lambda'\widetilde{R}_{a,1}(\omega), \\
     \widetilde{A}_\pm &= \widetilde{A}_\pm^0+\lambda'\widetilde{A}_\pm^1.
\end{align}
Replacing these in our system and solving perturbatively in powers of $\lambda'$ yields two algebraic systems. The first one, of zeroth order in $\lambda'$, is
\begin{equation}
    \begin{dcases}  \widetilde{A}_+^0e^{5i\alpha_+} + \widetilde{A}_-^0e^{5i\alpha_-} =  1\\
    \widetilde{A}_+^0e^{i\alpha_+} + \widetilde{A}_-^0e^{i\alpha_-} = 1\\
    \widetilde{A}_+^0e^{-i\alpha_+} + \widetilde{A}_-^0e^{-i\alpha_-} = \widetilde{R}_{a,0}(\omega)
    \end{dcases},
\end{equation}
whose solution is
\begin{equation}
   \widetilde{R}_{a,0}(\omega) = -2i\cos\left(\frac{\pi j}{2}\right)
\end{equation}
for all $a = \textsf{S}, \textsf{V}, \textsf{T}, \textsf{M}$ and
\begin{align}
     \widetilde{A}_+^0 &= \frac{ \widetilde{R}_ {a,0}(\omega)e^{i\alpha_-}-e^{-i\alpha_-}}{2 i \sin\left(\alpha_--\alpha_+\right)}, \\
     \widetilde{A}_-^0&= \frac{e^{-i\alpha_+}-\widetilde{R}_{a,0}(\omega)e^{i\alpha_+}}{2i\sin\left(\alpha_--\alpha_+\right)}.
\end{align}
The second system, of first order in $\lambda'$, is
\begin{equation}
    \begin{dcases}  \left(\widetilde{A}_+^1 + (\widetilde{\sigma}_+)_a\right)e^{5i\alpha_+} + \left(\widetilde{A}_-^1+ (\widetilde{\sigma}_-)_a\right)e^{5i\alpha_-} =    - L \omega T_{\text{H}}^{-1} \\
    \left(\widetilde{A}_+^1+ (\widetilde{\zeta}_+)_a\right)e^{i\alpha_+} + \left(\widetilde{A}_-^1 + (\widetilde{\zeta}_-)_a\right)e^{i\alpha_-} = 0 \\
    \left(\widetilde{A}_+^1+ (\widetilde{\zeta}_+)_a\right)e^{-i\alpha_+} + \left(\widetilde{A}_-^1 + (\widetilde{\zeta}_-)_a\right)e^{-i\alpha_-} = \widetilde{R}_{a,1}(\omega)
    \end{dcases},
\end{equation}
where $(\widetilde{\sigma}_\pm)_a$ and $(\widetilde{\zeta}_\pm)_a$ are respectively defined in equations~(5.88) and~(5.89) of~\cite{Moura:2024vhz}, under the redefinition~\eqref{23}. Solving this system for $\widetilde{R}_{a,1}(\omega)$ yields
\begin{equation}
   \widetilde{R}_{a,1}(\omega) =  -\frac{i}{2}\sec\left(\frac{\pi j}{2}\right)\left(2(\widetilde{\Sigma}_1)_a\cos\left(\pi j\right) - 2i(\widetilde{\Sigma}_2)_a\cos\left(\frac{\pi j}{2}\right) + (\widetilde{\Sigma}_1)_a + (\widetilde{\Sigma}_3)_a\right),
\end{equation}
where
\begin{align}
      (\widetilde{\Sigma}_1)_a &:= -\left((\widetilde{\zeta}_+)_ae^{i\alpha_+} +(\widetilde{\zeta}_-)_ae^{i\alpha_-} \right), \\
     (\widetilde{\Sigma}_2)_a &:= -\left((\widetilde{\zeta}_+)_ae^{-i\alpha_+} +(\widetilde{\zeta}_-)_ae^{-i\alpha_-} \right),\\
       (\widetilde{\Sigma}_3)_a &:=
      -\left((\widetilde{\sigma}_+)_ae^{5i\alpha_+} +(\widetilde{\sigma}_-)_ae^{5i\alpha_-}\right)-L\omega T_{\text{H}}^{-1}.
\end{align}

After some algebraic manipulation, we can rewrite $\widetilde{R}_a(\omega)$ as
\begin{equation}
   \widetilde{R}_a(\omega) = \widetilde{R}_{a,0}(\omega)\left(1 + \lambda'\delta \widetilde{R}_a(\omega)\right),
\end{equation}
where
\begin{equation}
    \delta \widetilde{R}_a(\omega) =  -\frac{i}{2\widetilde{R}_{a,0}(\omega)}\sec\left(\frac{\pi j}{2}\right)\left(2(\widetilde{\Sigma}_1)_a\cos\left(\pi j\right) - 2i(\widetilde{\Sigma}_2)_a\cos\left(\frac{\pi j}{2}\right) + (\widetilde{\Sigma}_1)_a + (\widetilde{\Sigma}_3)_a\right).
\end{equation}
Taking the limit $j\to 0$ for $a = \textsf{S},\textsf{T},\textsf{M}$ and $j\to 2$ for $a = \textsf{V}$ yields
\begin{align}
     \widetilde{R}_{a,0}(\omega) &= -2i, \hspace{3pt} a = \textsf{S},\textsf{T},\textsf{M}, \\
       \widetilde{R}_{\textsf{V},0}(\omega) &= +2i.
\end{align}
Furthermore, after some algebraic manipulation, we get
\begin{equation}
    \delta\widetilde{R}_a(\omega) = -\frac{(d-2)(d-4)}{16} \frac{\omega}{T_{\text{H}}} + \left(\frac{\omega}{T_{\text{H}}}\right)^{\frac{d-1}{d-2}}\left(\frac{d-3}{4\pi}\right)^{\frac{d-1}{d-2}}\widetilde{\Upsilon}_a,
\end{equation}
where we defined the constants
\begin{align}
      \widetilde{\Upsilon}_a &= -e^{-\frac{3\pi i}{2}\left(1 + \frac{1}{d-2}\right)} \frac{\pi ^2 2^{-\frac{1}{d-2}} (d-2)^{-\left(\frac{1}{d-2}+1\right)} (d-4) ((d-5) d+2)  \Gamma \left(\frac{1}{d-2}\right)}{(d-1) \Gamma \left(\frac{1}{2} \left(1+\frac{1}{d-2}\right)\right)^4}, \hspace{3pt} a = \textsf{S},\textsf{T},\textsf{V}, \\
     \widetilde{\Upsilon}_\textsf{M} &= - e^{-\frac{3\pi i}{2}\left(1 + \frac{1}{d-2}\right)}\frac{2 ^{-\frac{1}{d-2}}\pi ^2 (d-2)^{-\left(\frac{1}{d-2} + 4\right)} (d-4) (d-3) (d-1)^3   \Gamma \left(\frac{1}{d-2}\right)}{16  \Gamma \left(\frac{1}{2}\left(3 + \frac{1}{d-2}\right)\right)^4}.
\end{align}
%

\subsection{Computation of $\gamma_a(\omega)$}
\noindent

Finally, using equation~\eqref{30} yields
\begin{equation}
    \gamma_a(\omega) = 1 - R_a(\omega)\widetilde{R}_a(\omega) = \gamma_0(\omega)(1+\lambda'\delta\gamma_a(\omega)),
\end{equation}
where
\begin{align}
     \gamma_0(\omega) &= \frac{e^{\frac{\omega}{T_{\text{H}}}}-1}{3+e^{\frac{\omega}{T_{\text{H}}}}},
     \label{eq:einsteingrey}\\
    \delta\gamma_a(\omega) &= \frac{4}{1-e^{\frac{\omega}{T_{\text{H}}}}}\left(\delta\widetilde{R}_a(\omega) + \delta R_a(\omega)\right).
    \label{eq:correction}
\end{align}
After some algebraic manipulation, we can write
\begin{equation}
    \delta \gamma_a (\omega) =-\frac{4}{3+e^{\frac{\omega}{T_{\text{H}}}}}\left[ \frac{3}{16}(d-4)(d-2)\frac{\omega}{T_{\text{H}}} +\left(\frac{\omega}{T_{\text{H}}}\right)^{\frac{d-1}{d-2}}\left(\frac{d-3}{4\pi}\right)^{\frac{d-1}{d-2}}\varrho_a\right],
\end{equation}
where
\begin{align}
     \varrho_a &=  e^{-\frac{2\pi i}{d-2}}\frac{\pi ^{\frac{3}{2}}  (d-2)^{-\left(\frac{1}{d-2} + 1\right)} (d-4) ((d-5) d+2) \Gamma \left(\frac{1}{2(d-2)}\right)}{(d-1) \Gamma \left(\frac{1}{2} + \frac{1}{2(d-2)}\right)^3}\sin\left(\frac{\pi}{2(d-2)}\right), \hspace{3pt}  a = \textsf{S},\textsf{T},\textsf{V}, \\
     \varrho_\textsf{M} &= e^{-\frac{2\pi i}{d-2}}\frac{\pi ^{\frac{3}{2}} (d-2)^{-\frac{1}{d-2}} (d-4) (d-3) \Gamma \left(\frac{1}{2 (d-2)}\right)}{(d-1) \Gamma \left(\frac{1}{2}+\frac{1}{2(d-2)}\right)^3}\sin\left(\frac{\pi}{2(d-2)}\right).
\end{align}
%

\section{Conclusions}
\noindent

In this work, we have derived analytical formulae for the greybody factors of gravitational perturbations and of minimally coupled test scalar fields in the $d$-dimensional Gauss--Bonnet black hole, in the large black hole limit (see subsection~\ref{plbhl}). To gain analytical control, via standard WKB methods, over the scattering problem from which the greybody factors are extracted (see section~\ref{greybody}), we restricted our analysis to two limiting regimes: the eikonal limit and the highly damped limit.

\paragraph{The eikonal limit.} The eikonal limit, considered in section~\ref{eikonal}, assumes that the perturbations have large multipole number $\ell$ in their spherical harmonic decomposition. This assumption leads to a dramatic simplification of the master equation~\eqref{potential0}, making it amenable to WKB methods. Here, we have used the method developed in~\cite{Konoplya:2019hlu,Konoplya:2023moy} to obtain analytical expressions for the greybody factors in this regime. Our results were derived perturbatively to first order in $\lambda'$ (see equation~\eqref{lambda}). As expected, the zeroth-order term~\eqref{geik0} reproduces the known result for Einstein gravity, whereas the first-order correction admits a large-$\ell$ expansion~\eqref{dgeik}, whose two leading terms, proportional to $\ell$ and $\ell^{-1}$, we derive analytically. As in the Einstein gravity case, these terms depend strongly on the dimension $d$, for all types of gravitational perturbations as well as for the test scalar field.

\paragraph{The highly damped limit.}

The highly damped regime considered in Section~\ref{asymptotic} places no restriction on the parameters that define the perturbations. Instead, it assumes frequencies whose imaginary part is large in magnitude and much larger than the real part, that is, almost purely imaginary frequencies. The greybody factors obtained in this regime are therefore valid only in the corresponding region of the complex frequency plane. Our results were derived perturbatively to first order in $\lambda'$. As expected, the zeroth-order term~\eqref{eq:einsteingrey} reproduces the known Einstein gravity result of~\cite{Neitzke:2003mz}, while the first-order correction~\eqref{eq:correction} displays a similar structure. In particular, it has poles at the Einstein gravity quasinormal frequencies, implicitly defined by
\begin{equation}
    e^{\frac{\omega}{T_{\text{H}}}} + 3 = 0\,,
\end{equation}
and zeros at
\begin{equation}
    \omega = 2\pi i\times n\times T_{\text{H}}, \hspace{3pt} n \in \mathbb{Z}.
\end{equation}
This last property is not trivial: it arises only from the combination of the two reflection coefficients appearing in~\eqref{eq:correction}. Unlike the Einstein gravity result, the first order corrections display a strong dependence on the dimension $d$ for all types of gravitational perturbations as well as for the test scalar field.

Finally, we find identical results for all types of gravitational perturbations, and a different one for the scalar test field. This is somewhat remarkable, since the expansion of the master equation~\eqref{potential0} around $r = 0$ is one of the main inputs of the method employed here, and for vector perturbations this expansion differs from that of scalar and tensor perturbations, at both leading and subleading order in $\lambda'$. These differences nevertheless conspire to cancel, yielding exactly the same expression. Interestingly, the same degeneracy was observed in~\cite{Moura:2022gqm} for the quasinormal frequencies in the same highly damped regime, computed with the same monodromy method.

\noindent

\label{sec:conclusion}

\paragraph{Acknowledgements}
\noindent
This work has been supported by Funda\c c\~ao para a Ci\^encia e a Tecnologia under grants IT (UID/50008/2025), CAMGSD/IST-ID (UID/04459/2025) and project 2024.04456.CERN. JR is supported by the FCT-Portugal scholarship UI/BD/151499/2021 and by the CAMGSD scholarship BL197/2025-IST-ID.

\end{document}